\documentclass[epj]{svjour}

\usepackage{xcolor}
\usepackage{bm}
\usepackage{amsmath}
\usepackage{graphicx}
\usepackage{amssymb}
\usepackage{pifont}
\usepackage{array}
\usepackage{caption}

\begin{document}
\title{Microscopic bell-shape nonlocality: the case of proton scattering  off
  $^{40}$Ca at 200 MeV}
%
%

\author{G.~Blanchon\inst{1,2}\thanks{e-mail: guillaume.blanchon@cea.fr}
  \and H.~F. Arellano\inst{3,1}}

\institute{CEA,DAM,DIF, F-91297 Arpajon, France
\and
           Université Paris-Saclay, CEA, Laboratoire Matière sous
           Conditions Extrêmes, 91680 Bruyères-Le-Châtel, France 
\and
           Department of Physics - FCFM, University of Chile,
           Av. Blanco Encalada 2008, Santiago, RM 8370449, Chile           
          }

\abstract{
  This work is part of an ongoing effort to build microscopically-driven
  nonlocal optical potentials easily tracktable in scattering codes. 
  Based on the separable `$JvH$' structure proposed recently~\cite{arellano_22},
  where the potential can be cast as the product of a radial and 
  nonlocality form factors, we investigate its angular dependence
  in momentum space. We find that scattering observables have a 
  weak angular dependence between the momentum transfer 
  {$\bm q=\bm k-\bm k'$}, and {$\bm K=(\bm k+\bm k')/2$}.
  The study is focussed on proton elastic
  scattering off $^{40}$Ca at 200~MeV, where the $JvH$ structure
  is found to be inadequate. 
  We conclude that any improvement of the $JvH$ structure of the
  potential can be made to the lowest order in multipole
  expansions in $Kq-$representation of the potential.
}
\maketitle
\section{Introduction}
\label{sec::intro}

In the early sixties, Perey and Buck (PB) proposed a phenomenological nonlocal
optical potential for neutron scattering off spherical nuclei \cite{perey_62}.
This potential factorizes into a Gaussian nonlocal form factor and a
Woods-Saxon radial shape. The PB potential model has been further used
as a starting point for numerous phenomenological studies. In addition to its
extension to proton projectiles \cite{tian_15}, its imaginary component has
been made energy dependent \cite{lovell_17,jaghoub_18} and also 
dispersive~\cite{mahzoon_14,mahzoon_17,atkinson_20}.
\\

In a previous work \cite{arellano_22}, we have studied microscopic optical
model potentials based on a $g$-matrix approach \cite{arellano_89}, where
the potential can be reduced to lowest order to the so-called $JvH$ form which
relates directly to PB form factor. The work is done in momentum space, where
a simplification is obtained with the use of mean momentum $\mathbf{K}$ and
momentum transfer $\mathbf{q}$ instead of the usual $\mathbf{k}$ and
$\mathbf{k'}$ relative momenta. This $JvH$ potential is shown to reproduce
elastic observables for incident energies up to 40~MeV.
\\

In the present work we explore this microscopically-driven potential at
200~MeV incident energy. We start from microscopic potential based 
on a density-dependent infinite nuclear-matter model to represent the
\textit{in-medium} effective $NN$ interaction
\cite{arellano_95,arellano_07a,aguayo_08}, folded with proton and neutron
densities from Gogny D1S Hartree-Fock calculations
\cite{decharge_80}. The Brueckner-Hartree-Fock $g$-matrix is obtained using
AV18 bare interaction \cite{wiringa_95}. We restrict this work to proton
elastic scattering off $^{40}$Ca at 200~MeV.
In Sec.~\ref{sec::expansions}, we define two alternative representations:
$kk'$ and $Kq$, named after the coordinate system used. We truncate
the angular dependence in the $Kq$-representation. The truncated potential
is then expanded in $kk'$ which is suitable to solve the scattering
equations. The convergence of this multipolar expansion is assessed  in
Sec.~\ref{sec::convergence}. The underlying idea is to investigate whether
lower orders in $Kq$-expansion retains the leading physics at 200~MeV
incident energy. In Sec.~\ref{sec::jvh}, we assess the $JvH$ form for the
optical potential. Limitations of the $JvH$ form relative to its original
potential is illustrated in the context of elastic scattering. 

Results from this work show that the $Kq$-representation exhibits a weak
angular dependence between ${\bm q}$ and ${\bm K}$ allowing to focus on the term
participating to $J$, $v$ and $H$ terms. This microscopically-driven
potential should be tracktable for further phenomenology. Conclusions and
perspectives are drawn in Sec.~\ref{sec::conclusions}.

\section{Expansions}
\label{sec::expansions}
  
We consider the general structure of the optical potential in momentum space.
In all these approaches the optical potential for \textit{NA} elastic
scattering, ${\cal U}({\bm k}',{\bm k};E)$, can be cast in the 
form~\cite{arellano_22}
\begin{equation}
  \label{eq::ukk}
  {\cal U}({\bm k'},{\bm k};E) =
  {\cal U}_{c}({\bm k'},{\bm k}) +
  i{\bm\sigma}\cdot\hat{\bm n}\,
  {\cal U}_{so}({\bm k'},{\bm k})\;,
\end{equation}
with $\hat{\bm n}$ the unit vector perpendicular to the scattering plane given
by
\begin{equation}
  \hat{\bm n}=\frac{\bm k'\times\bm k}{|\bm k'\times\bm k|}\,,
\end{equation}
and ${\bm\sigma}$ the spin of the projectile. Here ${\cal U}_{c}$ and
${\cal U}_{so}$ represent the central and spin-orbit components of the
potential. From now on, dependences on the energy $E$ are implicit. One can
expand the potential in Eq.~\eqref{eq::ukk} in terms of relative momenta,
$k$ and $k'$, and the angle between ${\bm k}$ and ${\bm k}'$ expressed by
$u\!=\!\hat{\bm k}\cdot\hat{\bm k}'$, as  
\begin{equation}
 \label{eq::kkp-exp} 
 {\cal U}({\bm k'},{\bm k}) = \sum_{l=0}^{\infty}U_{l}(k,k')P_{l}(u).
\end{equation}
In the following, this expansion is referred to as $kk'-$expansion, where $l$
represents 
the orbital angular momentum \cite{joachain_75}. This is the form we use
to obtain the scattering observables using SWANLOP code \cite{arellano_21b}
that allows treatment of optical potential in momentum space as input.\\
Let us now define the following alternative variables,
\begin{align}
  {\bm K} =& \textstyle{\frac12}({\bm k} + {\bm k}'),\\
  {\bm q} =& {\bm k} - {\bm k}',\\
       w  =& \hat{\bm K}\cdot \hat{\bm q},
\end{align}
with ${\bm K}$ the mean momentum and ${\bm q}$ the momentum transfer. 
We observe that ${\bm k'}\times{\bm k}\!=\!{\bm K}\times{\bm q}$.
As done in Ref.~\cite{arellano_22}, we investigate the
structure of the potential in terms of ${\bm K}$ and ${\bm q}$.
Accordingly, we introduce the notation,
\begin{equation}
  \label{eq::notation}
  {\cal U}({\bm k'},{\bm k};E) = {U}({\bm K},{\bm q})
  = {U}_{c}(K,q,w) + i {\bm\sigma}\cdot({\bm K}\times{\bm q})\,
  {U}_{so}(K,q,w)\;.
\end{equation}
Now we consider the following expansions,
\begin{subequations}
\begin{align}
    {U}_{c}(K,q,w) &=
    \sum_{\substack{n=0\\ n\ even}}^{N} \,{U}^{(c)}_{n}(K,q) P_n(w)\;,
    \label{eq::pw0} \\
    |{\bm K}\times{\bm q}| {U}_{so}(K,q,w) &=
    Kq\sum_{\substack{n=0\\ n\ even}}^{N} \,{U}^{(so)}_{n+1}(K,q) P^{1}_{n+1}(w)\;,
    \label{eq::pw0-so}
\end{align}
\end{subequations}
for the central part and the spin-orbit potentials, respectively.
In principle $N\!\rightarrow\!\infty$, but in practice we choose an upper limit.
$P^{m}_{n}(w)$ is the associated Legendre polynomials. These expansions are
referred to as $Kq-$expansion. Restriction of summation to even $n$ values
in the central term comes from symmetry considerations. Note that the spin
orbit term is summed over odd values. Details of spin-orbit derivations
in Eq.~\eqref{eq::pw0-so} can be found in Appendix~\ref{sec::so}.

\section{Convergence}
\label{sec::convergence}

We now explore the ability of different $Kq-$truncations to reproduce the
full potential result in the context of the test case: proton elastic
scattering off $^{40}$Ca at 200~MeV. First of all, we verify that the exact
results are reproduced when using high values of $N$. \\

\begin{figure}[h!]
\centering
\includegraphics[scale=0.55,angle=00]{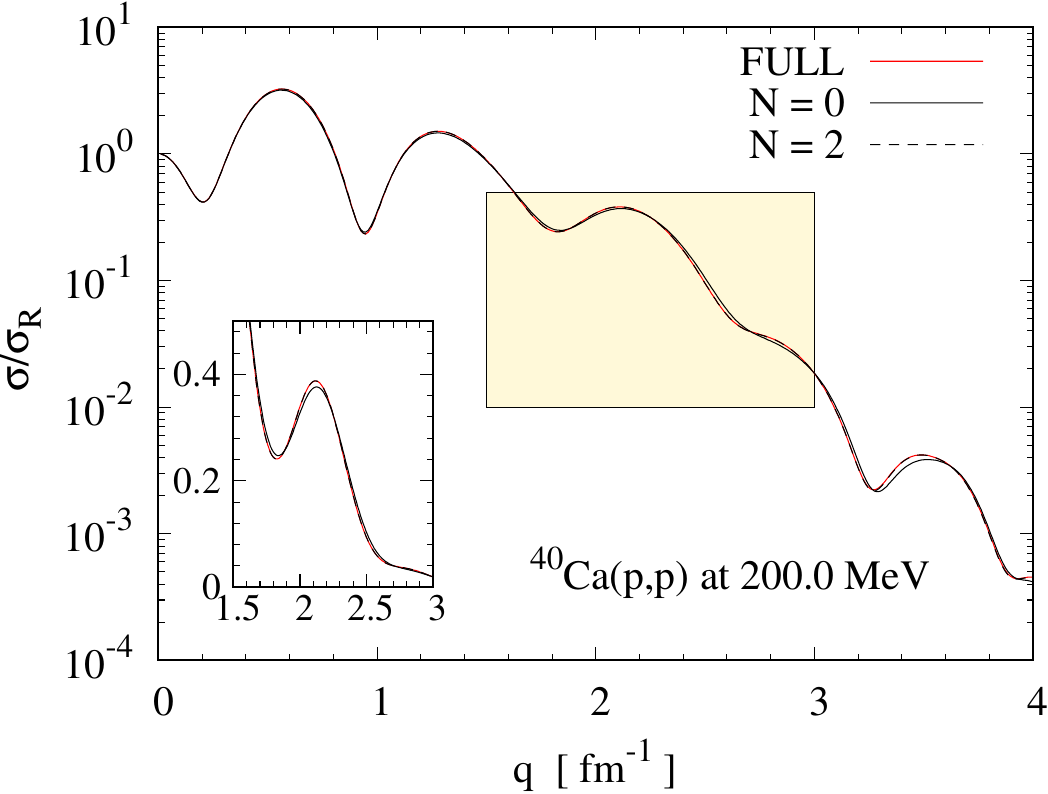}
\caption{Ratio-to-Rutherford differential cross section as a function of the
  momentum transfer for p~+~$^{40}$Ca at 200~MeV. 
Red curves correspond to the full potential, whereas 
	solid and dashed black curves represent results in $Kq-$repesentation
	truncated at $N\!=\!0$ and $2$, respectively.
	The inset shows a close-up in linear scale.
  }
\label{fig::fig1}
\end{figure}

In Fig.~\ref{fig::fig1}, we present the calculated differential cross sections
obtained from full potential as in Eq.~\eqref{eq::ukk}. We also show results
considering the cases $N\!=\!0$, and $N\!=\!2$ from Eqs.~\eqref{eq::pw0}
and~\eqref{eq::pw0-so}. We observe that the agreement between the full
calculation and that obtained for $N\!=\!0$, is already remarkable. The result
for $N\!=\!2$ 
overlaps perfectly to the eye with the full calculation. For the sake of scrutinity, we include
an inset where the differential cross section is plotted in linear scale, to
strengthen this statement. In Table~\ref{tab::reac}, we summarize the
calculated reaction cross sections considering the full representation of the
potential and its expressions up to $N\!=\!0$ and $2$. 
The difference between the full potential and the case of $N\!=\!0$ is about
0.4~\%, reaching agreement up to four significant figures in the case
of $N\!=\!2$.
These results validate considering the former case as starting point.\\
\begin{table}
\centering
\begin{tabular}{l|l}
\hline
\hline 
  Scheme & $\sigma_{R}$ [mb] \\
\hline 
  Full       &     526.5      \\
  $N\!=\!0$  &     528.4      \\
  $N\!=\!2$  &     526.5      \\
\hline
\hline
\end{tabular}
\caption{Reaction cross sections for p+$^{40}$Ca at 200 MeV for the full potential
  scattering calculation, the $N\!=\!0$ and $N\!=\!2$ truncations in the $Kq-$representation.}
\label{tab::reac}
\end{table}

\section{$JvH$ and beyond}
\label{sec::jvh}

We have shown that the optical potential in the $Kq-$representation 
exhibits a weak dependence on the angle between $\bm K$ and $\bm q$.
Actually, the truncation up to $N\!=\!0$ is reasonably good to retain
the leading features present in the original potential in
$kk'-$representation. This feature justifies to focus on the $N\!=\!0$
truncated expansion of the potential. 
In this case Eqs.~\eqref{eq::pw0} and~\eqref{eq::pw0-so} become
\begin{subequations}
\begin{align}
	{U}_{c}(K,q,w) &\approx {U}^{(c)}_{n=0}(K,q)\;,    \label{eq::pw1}\\
	{U}_{so}(K,q,w) &\approx   {U}^{(so)}_{n=1}(K,q)\, . \label{eq::pw1-so}
\end{align}
\end{subequations}
Note that both terms are independent of $w$, namely the angle between
${\bm K}$ and ${\bm q}$. 
Following Ref.\cite{arellano_22} we define
\begin{subequations}
\begin{align}
  V^{(c)}(K,q) &= \frac{U_{c}(K,q)}{U_{c}(K,0)},  \\
  H^{(c)}(K)   &= \frac{U_{c}(K,0)}{U_{c}(0,0)}.
\end{align}
\end{subequations}
The term $U_{c}(0,0)$ is directly related to the volume integral through, 
$J^{(c)}=(2\pi)^3 U_{c}(0,0)$. The same construction is applied to the
spin-orbit term. Below $E\!=\!40$~MeV, $V^{(c)}$ and $V^{(so)}$ are $K$-independent
\cite{arellano_22}. This is the key point to recover the Perey-Buck
separability between a nonlocal form factor and a radial shape.
Therefore we introduce the radial form factors,
\begin{align}
	V^{(c)}(K,q) &\approx V^{(c)}(0,q) \equiv v^{(c)}(q)\,, \label{vcq} \\
	V^{(so)}(K,q) &\approx V^{(so)}(0,q) \equiv v^{(so)}(q)\,. \label{vsoq} 
\end{align}
In this way one gets the $JvH$ form of the potential, 
\begin{equation}
  \label{eq::jvh}
  U^{JvH}(K,q) = \frac{J^{(c)}}{(2\pi)^3} \, v^{(c)}(q) \, H^{(c)}(K) 
  + i{\bm \sigma}\cdot ( {\bm K}\times{\bm q}) 
  \frac{J^{(so)}}{(2\pi)^3} \, v^{(so)}(q) \, H^{(so)}(K).
\end{equation}

\begin{figure}[h!]
\centering
\includegraphics[scale=0.4]{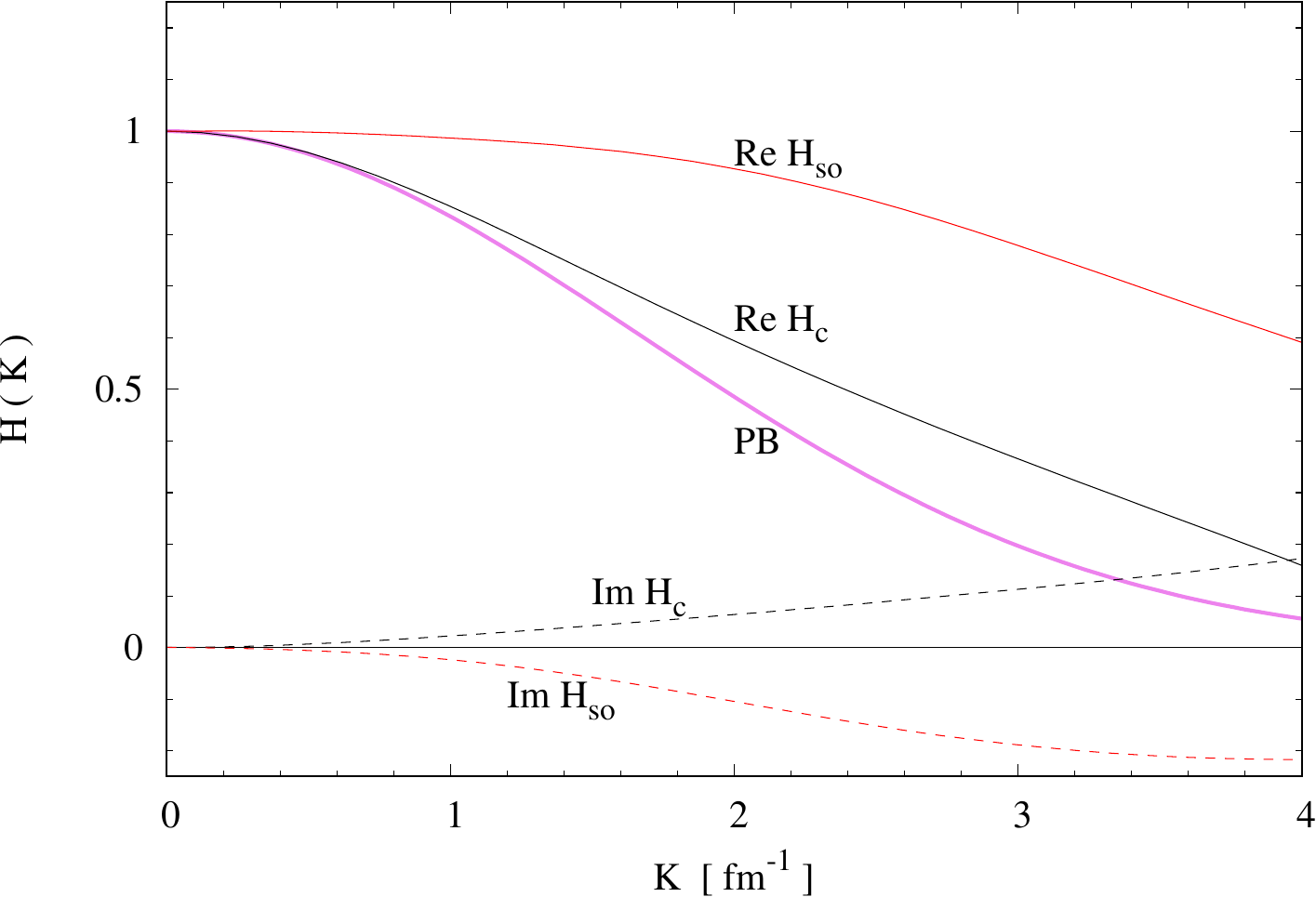}
\caption{Bell-shape nonlocality form factor $H(K)$ in Eq.~\eqref{eq::jvh} for
  p~+~$^{40}$Ca potential at 200~MeV. Results for central (c) and spin-orbit
  (so) compared with Perey-Buck nonlocality form factor with $\beta\!~=~\!0.85$~fm
  in Eq.~\eqref{eq::Hpb}.}
\label{fig::fig2}
\end{figure}

The behavior of the nonlocality of the microscopic optical potential
is illustrated in
Fig.~\ref{fig::fig2}, where we plot the nonlocality form factor $H$
as a function of $K$.
Black and red curves correspond to the central and spin-orbit
nonlocality form factors 
extracted from the microscopic optical potential, respectively.
Solid and dashed curves correspond to their real (Re) and imaginary (Im)
components, respectively.
Here we also include the nonlocality form factor
proposed by Perey-Buck \cite{perey_62}, namely
\begin{equation}
  \label{eq::Hpb}
  H^{PB}(K)=e^{-\beta^{2} K^{2}/4}\;,
\end{equation}
where the nonlocality parameter is taken as $\beta\!=\!0.85$~fm.
We observe two features worth mentioning. First, the microscopic
nonlocality form factor is complex, with a small imaginary contribution
at the origin but increasing in magnitude as $K$ increases.
Second, the curvature of $H$ at the origin is a measure of the
range of the nonlocality. Thus, the nonlocality of the spin-orbit
term is smaller than that in the central part. Interestingly,
the curvature of the central potential and PB appear comparable.

\begin{figure}[h!]
\centering
\includegraphics[scale=0.6,angle=-00]{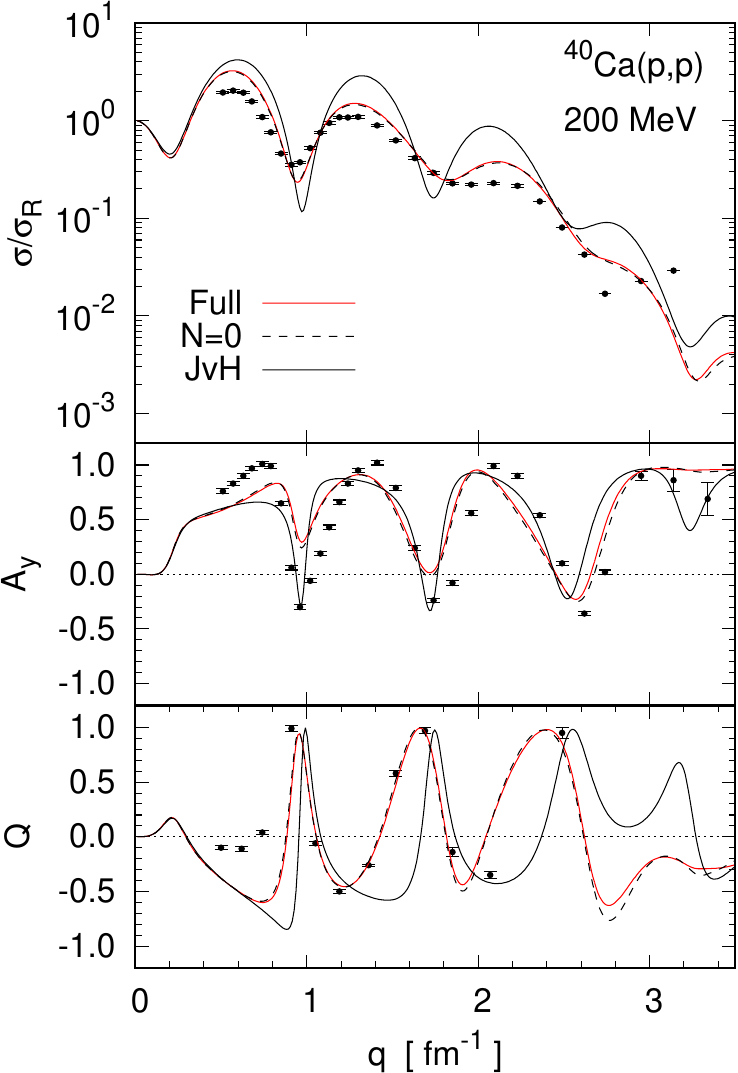}
\caption{Ratio-to-Rutherford differential cross section,
analyzing power and spin rotation as functions of the momentum transfer
	for p~+~$^{40}$Ca elastic scattering at 200~MeV. 
	Red solid curves are based on the full potential.
	Black dashed curves are based on its $Kq-$representation of order zero.
	Black solid curves represent results based on its $JvH$ factorization.
	The data are taken from Refs.~\cite{hutcheon_88} and 
  \cite{stephenson_85}.
  }
\label{fig::fig3}
\end{figure}

In Fig.~\ref{fig::fig3}, we show results for the
ratio-to-Rutherford differential cross sections $\sigma/\sigma_R$,
analyzing power $A_y$ and spin-rotation $Q$ as functions of the
momentum transfer $q$. The data are taken from Refs.~\cite{hutcheon_88} and \cite{stephenson_85}. 
Red solid curves represent results based on the full potential. 
Black dashed curves correspond to $Kq-$representation considering 
$N\!=\!0$.
These results illustrate the close correspondence between the full potential
and its $Kq-$representation up to $N\!=\!0$, whereas clear differences
are evidenced when compared to the $JvH$ form.
Thus, the origin of these differences can be traced back to the use
of 
Eqs.~\eqref{vcq} and \eqref{vsoq}, where the $K$ dependence is neglected.
This observation indicates that any improvement of the $JvH$ factorization
would require corrections in $K$ at this level. Work along this line is
underway.

\section{Conclusions}
\label{sec::conclusions}

The main result of this work is that $Kq-$representation of the optical model potential
retains the main physics at its zero-th multipolar order. 
This feature is unveiled from a microscopic optical model potentials
based on realistic bare nucleon-nucleon potentials.
Although the $JvH$ form has been validated at low incident energies
in Ref.~\cite{arellano_22}, its limitations above 100 MeV can be
corrected after a close scrutiny of the
$K$ dependance in Eqs.~\eqref{vcq} and \eqref{vsoq}.
The long term aim of this effort is to provide guidance to phenomenomogical
constructions of optical potentials resulting from detailed microscopic
approaches.

\appendix

\section{Spin-orbit term}
\label{sec::so}
We describe in details how to get the expansion for the spin-orbit contribution in
Eq.~\eqref{eq::pw0}. We start from the usual expression of the spin-orbit,
\[ 
\hat U = \hat U_{so}\,{\bm\ell}\cdot{\bm\sigma}\;.
\]
Then we get the $kk'-$representation of the potential,
\begin{align}
\langle{\bm k'}|
\hat U
|{\bm k}\rangle
&= 
\langle{\bm k'}|
  \hat U_{so}\,{\bm\ell}\cdot{\bm\sigma}
|{\bm k}\rangle \\
&= 
\langle{\bm k'}|
  \hat U_{so}\,i\hbar({\bm r}\times{\bm p})\cdot{\bm\sigma}
|{\bm k}\rangle \\
&= 
i\hbar
\langle{\bm k'}|
  \hat U_{so}\,\nabla_k \times {\bm k} \,
|{\bm k}\rangle 
  \cdot{\bm\sigma}.
\end{align}
Denote with $\partial_i$ the partial derivative with
respect to the $i$-th component of momentum ${\bm k}$,
we get
\begin{align}
\langle{\bm k'}|
\hat U
|{\bm k}\rangle
&= 
i\hbar \sum_{n}
\epsilon_{imn} \partial_i 
  \left (
  k_m 
\langle{\bm k'}|
  \hat U_{so}\,
  |{\bm k}\rangle
  \right )
  {\sigma}_n \\
&= 
i\hbar \sum_{n}
\epsilon_{imn} 
  k_m 
  \partial_i 
  \left (
\langle{\bm k'}|
  \hat U_{so}\,
  |{\bm k}\rangle
  \right )
  {\sigma}_n \\
&= 
i\hbar \sum_{n}
\epsilon_{nim} 
  {\sigma}_n 
  \partial_i 
  \left (
\langle{\bm k'}|
  \hat U_{so}\,
  |{\bm k}\rangle
  \right )
  k_m   \\
&= 
i\hbar
  {\bm\sigma} 
  \cdot
  \left (
  \nabla
\langle{\bm k'}|
  \hat U_{so}\,
  |{\bm k}\rangle
  \right )
  \times
   {\bm k}.
\end{align}
Let us define,
\[
\langle{\bm k'}|
  \hat U_{so}\,
  |{\bm k}\rangle =
  \sum_l U_l(k',k) P_l(\hat k\cdot\hat k'),
\]
with ${\bm k'}$ the azimuthal, then the non vanishing
contribution above comes from angular variations in 
$\cos\theta=\hat k\cdot\hat k'$.
Hence
\begin{align}
\langle{\bm k'}|
\hat U
|{\bm k}\rangle
&= 
i\hbar
  {\bm\sigma} 
  \cdot
  \left (
  \nabla
\langle{\bm k'}|
  \hat U_{so}\,
  |{\bm k}\rangle
  \right )
  \times
   {\bm k}   \\
&= 
i\hbar
  {\bm\sigma} 
  \cdot
  \left (
  \frac{\partial}{k\,\partial\theta}
  \left [
  \sum_l U_l(k',k) P_l(\hat k\cdot\hat k')
  \right ]
\hat\theta
  \right )
  \times
   {\bm k}   \\
&= 
i\hbar
  {\bm\sigma}\cdot(\hat\theta\times\hat k) 
  \left (
  \frac{\partial}{\partial\theta}
  \left [
  \sum_l U_l(k',k) P_l(\hat k\cdot\hat k')
  \right ] 
  \right ) \\
&= 
i\hbar
  {\bm\sigma}\cdot\hat n
  \left (
    \sum_l U_l(k',k) P_l^{1}(\hat k\cdot\hat k')
  \right ), 
\end{align}
where we identify
\[
  \hat n=\frac{{\bm k'}\times{\bm k}}{|{\bm k'}\times{\bm k}|},
\]
and
\[
  P_l^{1}(\cos\theta) = -\sin\theta \, \frac{dP_l(u)}{du},
\]
with $P_{l}^{m}(x)$ the associated Legendre polynomials.

\bibliography{references}

\end{document}